\documentclass[twocolumn,showpacs,preprintnumbers,amsmath,amssymb]{revtex4-1}

\usepackage{graphicx}
\usepackage{dcolumn}
\usepackage{bm}

\def\bequ{\begin{equation}}
\def\eequ{\end{equation}}
\def\be{\begin{equation}}
\def\ee{\end{equation}}

\begin{document}


\title{Hawking radiation for a Proca field in $D$ dimensions \uppercase\expandafter{\romannumeral2}: \\charged field in a brane charged black hole}

\author{Mengjie Wang}
\email{mengjie.wang@ua.pt}
\author{Marco O. P. Sampaio}
 \email{msampaio@ua.pt}
 \author{Carlos Herdeiro}
\email{herdeiro@ua.pt}
\affiliation{\vspace{2mm}Departamento de F\'\i sica da Universidade de Aveiro and I3N \\
Campus de Santiago, 3810-183 Aveiro, Portugal \vspace{1mm}}%

\date{\today}

\begin{abstract}
We generalise our first analysis of the wave equation for a massive vector boson in the background of a D-dimensional Schwarzschild black hole~\cite{HSW:2012}, by adding charge both to the field and the black hole, on the 3+1 dimensional Standard Model brane.
A detailed numerical study is performed to obtain the transmission factor for the coupled (as well as decoupled) system of equations describing the Proca field modes, varying the angular momentum number, mass, charge and space-time dimensions. A qualitatively new feature arising from the introduction of charge is the appearance of \textit{superradiant modes}, which we investigate.  We then compute the Hawking fluxes, and analyse the effect of the charge. In particular we observe an inverted charge splitting effect for small energies and for two or more extra dimensions. For neutral fields, we compare the emission of massive particles with spins up to one and also compare the Proca bulk-to-brane ratio of energy emission, showing that, as for a scalar field, most of the energy is emitted on the brane.

\end{abstract}

\maketitle


\section{Introduction}

The possible existence of additional space-like dimensions in nature has been discussed for a century, at least since the works by Nordstr\"om, Kaluza and Klein. During the last four decades, moreover, the naturalness of extra dimensions within Supergravity and String theory made it a main stream topic within high energy theoretical physics. At the end of the last century, this research led to models that, aiming at solving the hierarchy problem, predicted that the extra dimensions could be very large (or even infinite), as compared to the traditional Planck scale. Within such scenario, the true fundamental Planck scale could be as low as the TeV scale~\cite{Antoniadis:1990ew,Arkani-Hamed:1998rs,Antoniadis:1998ig,Arkani-Hamed:1998nn} so that strong gravity effects could be visible in realistic man made particle accelerators. In  particular, miniature black holes would be produced in particle collisions and Hawking radiation would be the main observable signature~\cite{Banks:1999gd,Dimopoulos:2001hw,Giddings:2001bu}. This motivation led to an intensive study of Hawking radiation from higher-dimensional black holes~\cite{Ida:2002ez,Harris:2003eg,Harris:2005jx,Ida:2005ax,Duffy:2005ns,Casals:2005sa,Cardoso:2005vb,Cardoso:2005mh,Ida:2006tf,Casals:2006xp,Casals:2008pq,Sampaio:2009ra,Sampaio:2009tp}.

Though so far there is no experimental evidence for the existence of such extra dimensions or strong gravity events in the currently probed energy frontier at short distances, the bounds are not yet as strong as one might wish~\cite{CMS:2012yf,ATLAS-CONF-2011-065,ATLAS-CONF-2011-068,Cardoso:2012qm,Park:2012fe,Gingrich:2012vs}. Thus, it is still of interest to understand how such events should reveal themselves, as to impose the best possible bounds~\cite{Frost:2009cf,Dai:2007ki}, especially for future high energy experiments, where a semi-classical treatment of the Hawking evaporation becomes a better approximation. Such an approximation holds in the highly trans-planckian regime, otherwise quantum gravity corrections become important~\cite{Agullo:2006iv,Agullo:2006um,Jacobson:1993hn}.

One of the Hawking radiation channels that has not been properly addressed in the literature is that of massive vector bosons, both uncharged and charged, to describe the emission of $Z$ and $W^{\pm}$ particles of the Standard Model (SM). The basic difficulty is that the Proca equations do not decouple even in a spherically symmetric black hole background. To bridge this gap, we have recently constructed a numerical strategy to investigate the transmission factors, as well as the Hawking radiation flux, for a neutral Proca field in a $D$ dimensional Schwarzschild space-time~\cite{HSW:2012}. In the present paper we close this study~\cite{Sampaio:2009ra,Sampaio:2009tp} by considering the Proca field on a brane and including charge for both the background and the spin 1 field, as to model the charged vector bosons of the SM.

A qualitatively new feature that occurs when both background and Proca field charges are included is the existence of superradiant modes.
For a bosonic field mode with frequency $\omega$ and azimuthal quantum number $m$, it is expected to observe superradiant scattering in the background of a rotating black hole with horizon angular velocity $\Omega_H$ when $\omega<m\Omega_H$. These modes are amplified by extracting rotational energy and angular momentum from the black hole. A similar amplification effect can be observed for a charged bosonic field mode, with frequency $\omega$ and charge $q$, when impinging in a charged black hole with electric potential at the horizon $\Phi_H$, when $\omega<q\Phi_H$. Such mode extracts Coulomb energy, as well as charge, from the charged black hole. In a rotating background, the Proca field equation variables are not known to separate, which presents an extra difficulty added to the non decoupling of the modes, making it difficult to study exactly the superradiance phenomenon - see~\cite{paniPRL,paniPRD} for a recent study in the slow rotation approximation. Thus, as a second motivation to our work, the charged black hole background, due to the spherical symmetry, yields a setup where superradiance of a massive spin 1 field can be studied without any approximation, albeit numerically. Such analysis will be performed herein.

The structure of this paper is organised as follows. In Section \ref{sec:q0Proca} we introduce the background geometry which will be explored in this paper. In Section \ref{sec3} we present the equation of motion for a charged Proca field on the charged brane; we discuss the boundary conditions and its conversion to a first order system adequate for numerical integration in Section \ref{sec:nearhorizon}. The numerical results for the transmission factor and the associated Hawking fluxes are presented in Section \ref{sec_results} and we close with a discussion and final remarks. To keep the main part of this paper compact and clear, some technical relations are left to an Appendix.

We shall use the particle physics signature convention $\left(+---\right)$ in this paper.

\section{Proca field and background}
\label{sec:q0Proca}
We consider a Proca field $W^{\mu}$ with mass $M$ and charge $q$, on a charged brane, which describes the SM $W$ particle. The Lagrangian for such field in the SM of particle physics (after electroweak symmetry breaking in unitary gauge) is
\begin{equation}
\mathcal{L}= -\dfrac{1}{2}W^\dagger_{\mu\nu}W^{\mu\nu}+M^2W_\mu^\dagger W^\mu-iqW_\mu^\dagger W_\nu F^{\mu\nu} \; ,\label{Procalag}
\end{equation}
where $W_{\mu\nu}=\mathcal{D}_\mu W_\nu-\mathcal{D}_\nu W_\mu$, $\mathcal{D}_\mu\equiv \partial_\mu-i q A_\mu$ and we have included the coupling of the $W$ field to the electromagnetic field strength tensor $F_{\mu\nu}=\partial_\mu A_\nu-\partial_\nu A_\mu$ as allowed by gauge invariance in the SM. The electromagnetic potential is denoted as $A_\mu$. We consider the following background brane black hole geometry
\begin{equation}
ds^2=V(r)dt^2-\dfrac{1}{V(r)}dr^2-r^2(d\theta^2+\sin^2\theta d\varphi^2) \ ,
\end{equation}
with metric function
\begin{equation}
V(r)=1-\dfrac{\mu}{r^{n-1}}+\dfrac{Q^2}{r^2}\;,\nonumber
\end{equation}
where $\mu$ and $Q$ are the parameters related with black hole bulk mass and brane charge. For numerical convenience, we choose units such that the outer horizon radius is $r_H=1$, i.e. $\mu=1+Q^2$ at the outer horizon.

\section{System of radial equations}
\label{sec3}
Due to the spherical symmetry of the background, we can obtain the equations of motion, using the gauge-invariant method of~\cite{Kodama:2000fa}, following the same procedure as we have done in~\cite{HSW:2012}. The form of the equations is very similar to that obtained therein; we therefore present here only a summary of the various radial equations (after Fourier transforming with respect to time and factoring out the angular harmonics).

\subsection{Massive $W^\mu$}
There are two sub-cases which have to be dealt with separately. Modes with $\ell \neq 0$, where $\ell$ is the usual angular momentum quantum number, are described by two coupled radial fields ($\psi,\chi$), corresponding to two propagating degrees of freedom and a decoupled transverse field $\Upsilon$ \footnote{The corresponding field combinations are $W_t-\mathcal{D}_t\Phi$, $W_r-\mathcal{D}_r\Phi$ and $W_i^T$, where the angular components $W_i=\partial_i\Phi+W_i^T$ are built from a scalar and transverse vector, with respect to the 2-sphere, respectively.}. The radial equations for the coupled fields are
\begin{eqnarray}
&&\left[V^2\dfrac{d}{dr}\left(r^2\dfrac{d}{dr}\right)+\left(\omega r -qQ\right)^2-V\left(\ell(\ell+1)+M^2r^2\right)\right]\psi\nonumber  \\
&&+\left[2iV\omega r-ir\left(\omega r-qQ\right)V'\right]\chi=0 \ ; \nonumber \\
&&\left[V^2r^2\dfrac{d^2}{dr^2}+\left(\omega r -qQ\right)^2-V\left(\ell(\ell+1)+M^2r^2\right)\right]\chi \nonumber \\ && +\left[2iqQV-i r\left(\omega r-qQ\right)V'\right]\psi=0 \; \;, \; \; \; \; \label{coupledequations}
\end{eqnarray}
whereas the decoupled equation for the transverse mode is
\begin{eqnarray}
&&\left[V\dfrac{d}{dr}\left(V\dfrac{d}{dr}\right)+\left(\omega -\tfrac{qQ}{r}\right)^2-\left(\tfrac{\ell(\ell+1)}{r^2}+M^2\right)V\right] \Upsilon \nonumber=0\,. \\
&& \label{TmodeEq}
\end{eqnarray}
Here $\ell$ starts from one, and $\omega$ is the frequency of the modes.
For the special case $\ell=0$, there is a single mode, as expected, and the two other modes are non-dynamical
\footnote{The scalar $\Phi$ does not appear in the equations of motion and $W_r$ can be eliminated in favour of $W_t$ or vice-versa}
\begin{multline}\left[\dfrac{d}{dr}\left(\dfrac{Vr^4}{(\omega r-qQ)^2-Vr^2M^2}\dfrac{d}{dr}\right)+\dfrac{r^2}{V} \right. \nonumber \\ \left. -\dfrac{qQr^2\left((\omega r-qQ) r V'-2qQV\right)}{\left[(\omega r-qQ)^2-Vr^2M^2\right]^2}\right]\psi^{(0)}=0 \ .
\end{multline}

\subsection{Massless $W^\mu$}
A special case, whose modes must match two of the modes in the previous subsection when $M\rightarrow 0$, is the exactly massless theory. This is an interesting case for various reasons. Firstly it provides a way to check the results of solving the coupled system in the previous section in a special limit. Secondly, it must provide accurate results for two of the degrees of freedom of the Proca field at high energies. Finally, the radial equations for this special case are all decoupled, so the definition of transmission factors is straightforward. Again this works as a numeric check of the results for the transmission factors of the coupled fields when $M\neq 0$.

In this special case, the $\ell=0$ mode is non-dynamical, i.e. its wave equations are satisfied identically. For $\ell \neq 0$, there is a transverse vector mode which is obtained by setting $M=0$ in Eq.~\eqref{TmodeEq}, and another mode described by $\chi$ which satisfies
\begin{multline}
\left[r^2V\dfrac{d}{dr}\left(V\dfrac{d}{dr}\right)-\dfrac{2qQV^2r}{(\omega r -qQ)}\dfrac{d}{dr}\right.\\\left.\phantom{\dfrac{V^2}{V^2}}+\left(\omega r  - qQ\right)^2-\ell(\ell+1)V\right]\chi =0 \ ,
\end{multline}
whereas $\psi=i r V d_r\chi/(\omega r-qQ)$ is non-dynamical.
We would like to anticipate that despite the apparent different form for these two modes, $\chi$ and $\Upsilon$, we shall find they give the same numerical results for the transmission factor, in all dimensions.

Since decoupled radial equations have been extensively studied in the literature, and a similar coupled system has been studied in detail in~\cite{HSW:2012}, in the remainder we will only review some of the details of the numerical set up to solve the coupled system.

\section{Boundary conditions and first order system}\label{sec:nearhorizon}
To determine the transmission factors, we need to integrate the radial equations from the horizon to the far away region with ingoing boundary conditions. The standard procedure is to find a series expansion of the solution near the horizon, which can be used to initialise the solution (we do so at $r=1.001$). Focusing on the coupled system $\left\{\psi,\chi\right\}$, if we define $y=r-1$, Eqs.~\eqref{coupledequations} become
\begin{eqnarray}
\left[A(r)\dfrac{d^2}{dy^2}+B(r)\dfrac{d}{dy}+C(r)\right]\psi+ E(r)\chi&=&0\label{systerm1}\ ,\\
\left[\tilde A(r)\dfrac{d^2}{dy^2}+\tilde
B(r)\dfrac{d}{dy}+\tilde C(r)\right]\chi+\tilde E(r)\psi&=&0\label{systerm2}\ ,
\end{eqnarray}
where the polynomials $A,B,C,E,\tilde{A},\tilde{B},\tilde{C},\tilde{E}$ are defined in the Appendix. Making use of Frobenius' method to expand $\psi$ and $\chi$, we insert the following expansions into Eqs.~\eqref{systerm1} and~\eqref{systerm2}
\begin{equation}
\psi=y^\rho\sum^{\infty}_{j=0}{\mu_jy^j}\label{defpsi} \;\; , \;\; \chi=y^\rho\sum^{\infty}_{j=0}{\nu_jy^j} \;\; , \;\; \rho=\tfrac{-i(\omega-qQ)}{(n-1)+(n-3)Q^2}
\end{equation}
where the sign of $\rho$ was chosen to impose an ingoing boundary condition, and we obtain the recurrence relations~\eqref{recurone} for the coefficients $\mu_j$ and $\nu_j$ found in the Appendix. A general solution close to the horizon can be parametrised by two free coefficients $\nu_0$ and $\nu_1$.

Similarly, to understand the asymptotic behaviour of the waves at infinity we now expand $\psi$ and $\chi$ as
\begin{equation}
\psi=e^{\beta r}r^{p}\sum_{j=0}\dfrac{a_j}{r^j}\label{psifar} \; , \qquad \chi=e^{\beta r}r^{p}\sum_{j=0}\dfrac{b_j}{r^j} \ ,
\end{equation}
which after insertion into Eqs.~\eqref{coupledequations} yield
\begin{equation}
\beta = \pm ik\; , \qquad p = \pm i\varphi \; ,
\end{equation}
where $\varphi=\delta_{n,2}(\omega^2+k^2)(1+Q^2)/(2k)-qQ\omega/k$, and $k=\sqrt{\omega^2-M^2}$. Thus one can show that  asymptotically
\begin{equation}
\psi \rightarrow \left(a_0^++\dfrac{a_1^+}{r}+\ldots\right)e^{i\Phi}+\left(a_0^-+\dfrac{a_1^-}{r}+\ldots\right)e^{-i\Phi} \; , \label{asymptoticpsi}
\end{equation}
\begin{eqnarray}
\chi \rightarrow &&
\left[\left(-\frac{k}{\omega}+\dfrac{c^+}{r}\right)a_0^++\ldots\right]e^{i\Phi} \nonumber \\
&&+\left[\left(\frac{k}{\omega}+\dfrac{c^-}{r}\right)a_0^-+\ldots\right]e^{-i\Phi} \; ,\label{asymptoticpchi}
\end{eqnarray}
where $\Phi\equiv kr+\varphi \log r$ and $c^\pm$ is defined in the Appendix, Eq.~\eqref{cpm}.
Thus, as expected, each field is a combination of ingoing and outgoing waves at infinity.
Asymptotically, the solution is parametrised by four independent coefficients $\left\{a_0^\pm,a_1^\pm\right\}$, two for each independent mode in the coupled system. In the same way as in~\cite{HSW:2012}, one can define a first order system of ODEs containing four radial functions $\left\{\chi^\pm,\psi^\pm\right\}$ which coincide with such coefficients at infinity, allowing for an easy extraction of the wave amplitudes. Our target system which will be solved numerically in the remainder is then
\begin{equation}\label{eq:ODEcoupled}
\dfrac{d\mathbf{\Psi}}{dr}=\mathbf{T}^{-1}\left(\mathbf{X}\mathbf{T}-\dfrac{d\mathbf{T}}{dr}\right) \mathbf{\Psi} \ ,
\end{equation}
with $\mathbf{\Psi}^T=(\psi_{+},\psi_{-},\chi_{+},\chi_{-})$. The definition of the matrices $\mathbf{X}$ and $\mathbf{T}$, and how they relate with~\eqref{systerm1} and~\eqref{systerm2} are found in the Appendix.

\section{Hawking fluxes}
\label{sec_results}
We shall now calculate the transmission factor for the coupled system as well as the Hawking fluxes generated from all the modes. In~\cite{HSW:2012} we have found a conserved flux from the energy momentum tensor which was simple enough to extract the transmission factors. However, in the present case, the energy-momentum tensor is more difficult to deal with so we use instead the conserved electric current which is naturally defined for this charged field. One can show that such a current is given by
\begin{equation}
\mathcal{J}^{\alpha}=W^{\dagger\alpha\mu}W_{\mu}+\dfrac{1}{\sqrt{-g}}\partial_{\beta}\left(\sqrt{-g}W^{\dagger\beta}W^{\alpha}\right)-c.c. \;\label{current}
\end{equation}
The radial flux at $r$ fixed is obtained by integrating the $\alpha=r$ component on the sphere. We note that only the first term (denoted $\mathcal{J}^{\alpha}_{\uppercase\expandafter{\romannumeral1}}$) contributes, since the second term becomes a total derivative on the sphere.

The contribution for the flux for the coupled fields at infinity, is then found by simplifying the radial component of Eq.~\eqref{current} using the equations of motion, and inserting the far away expansion at infinity
\begin{eqnarray}\label{eq:currentInf}
\mathcal{J}^{r, \infty}_{\uppercase\expandafter{\romannumeral1}-couple}&& =|y_0^-|^2-|y_0^+|^2+|y_1^-|^2-|y_1^+|^2
\nonumber
\\
&& \equiv(\mathbf{y}^-)^\dagger \mathbf{y}^--(\mathbf{y}^+)^\dagger \mathbf{y}^+ \; ,
\end{eqnarray}
where $y_i^s(s=\pm; i=0,1)$ are linear combinations of the asymptotic coefficients $a_i^s$ given in the Appendix, Eq.~\eqref{eq:yplus}.
Using the reflection matrix $(\mathbf{R})$ defined in~\cite{HSW:2012}, we obtain
\begin{equation}\label{eq:TF1}
\mathcal{J}^{r, \infty}_{\uppercase\expandafter{\romannumeral1}-couple}=(\mathbf{y}^-)^\dagger\left(\mathbf{1}-\mathbf{R}^\dagger\mathbf{R}\right) \mathbf{y}^-\equiv (\mathbf{y}^-)^\dagger\mathbf{T}\, \mathbf{y}^-  \; ,
\end{equation}
where we have defined a (hermitian) transmission matrix $\mathbf{T}$, which can be diagonalised such that the decoupled asymptotic fields are found.

Following the same procedure, one can also calculate the electric current at the horizon
\begin{equation}\label{eq:currentH}
\mathcal{J}^{r, H}_{\uppercase\expandafter{\romannumeral1}-couple}=\dfrac{1}{\omega-qQ}\left(\mathbf{h^-}\right)^\dagger\mathbf{h^-} \; ,
\end{equation}
where the $h^-_i$ coefficients are linear combinations of the two independent $\nu_i$ coefficients ($i=0,1$), given in the Appendix, Eqs.~\eqref{eq:hminus}. Eq.~\eqref{eq:currentH} shows the important point that the current can be positive or negative. This is expected, because for the boson field, the electric coupling can trigger superradiance.

Furthermore, from the conservation law of the electric current, one can find an alternative expression for the transmission matrix. Using the scattering matrix $(\mathbf{S}^{--})$ defined in~\cite{HSW:2012}, we find
\begin{equation}\label{eq:TF2}
\mathbf{T}=\dfrac{1}{\omega-qQ}(\mathbf{S}^{--}\mathbf{S}^{\dagger--})^{-1} \; .
\end{equation}

Once we have obtained the transmission factors, the number and energy fluxes are given by
\begin{equation}\label{eq:HawkFlux}
\dfrac{d\left\{N,E\right\}}{dt d\omega}=\dfrac{1}{2\pi}\sum_{\ell,\zeta} \dfrac{(2\ell+1)\left\{1,\omega \right\}}{\exp((\omega-qQ)/T_H)-1} \mathbb{T}_{\ell,\zeta} \;,
\end{equation}
where $\zeta$ labels the mode and  $T_H$ is the Hawking temperature which, in our units, is
\begin{eqnarray}
T_H=\dfrac{(n-1)+(n-3)Q^2}{4\pi}\;.\label{hawkingtemperature}
\end{eqnarray}

\section{Numerical Results}
\label{results}
We now present a selection of numerical results for the transmission factor and the corresponding Hawking fluxes. In order to integrate the decoupled and coupled radial equations, we wrote independent codes in \textsc{mathematica}7 and in \textsc{c++}, finding agreement between the two codes. Using them we have generated a set of figures that we now describe.

\begin{figure*}
\begin{center}
\begin{tabular}{ccc}
\includegraphics[clip=true,width=0.33\textwidth]{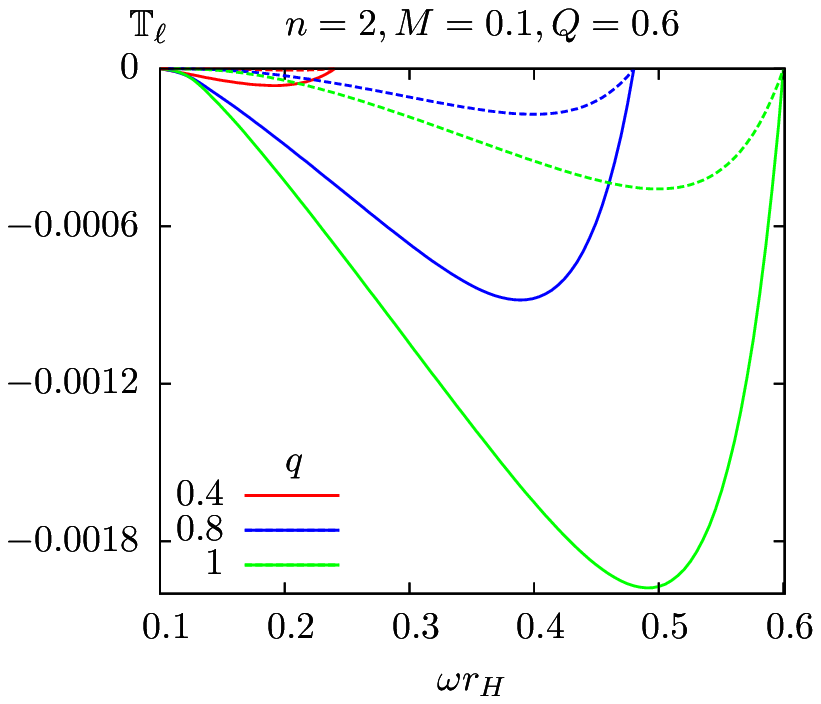}
\includegraphics[clip=false,width=0.312\textwidth]{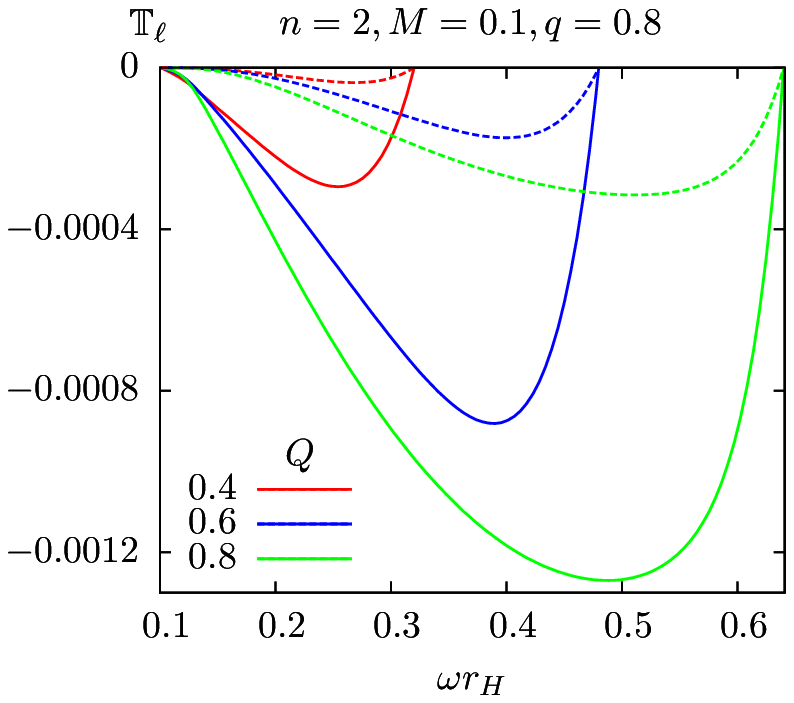} \hspace{2mm}
\includegraphics[clip=true,width=0.33\textwidth]{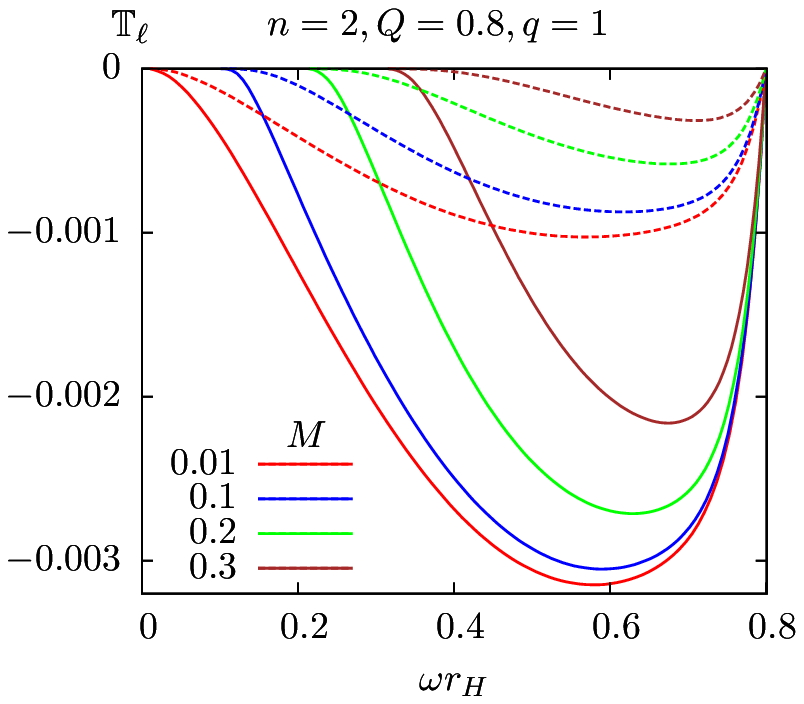}\\
\includegraphics[clip=true,width=0.33\textwidth]{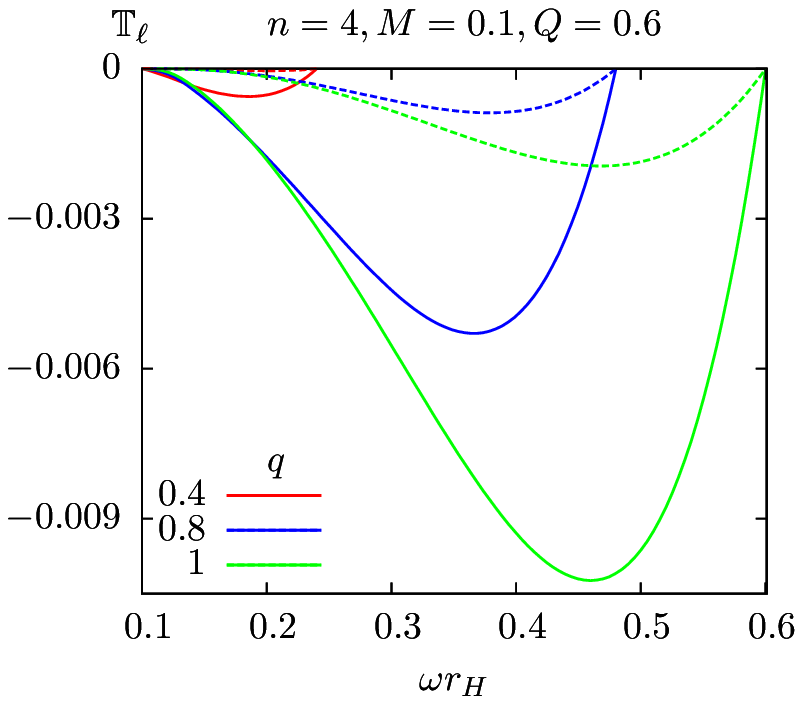}
\includegraphics[clip=false,width=0.312\textwidth]{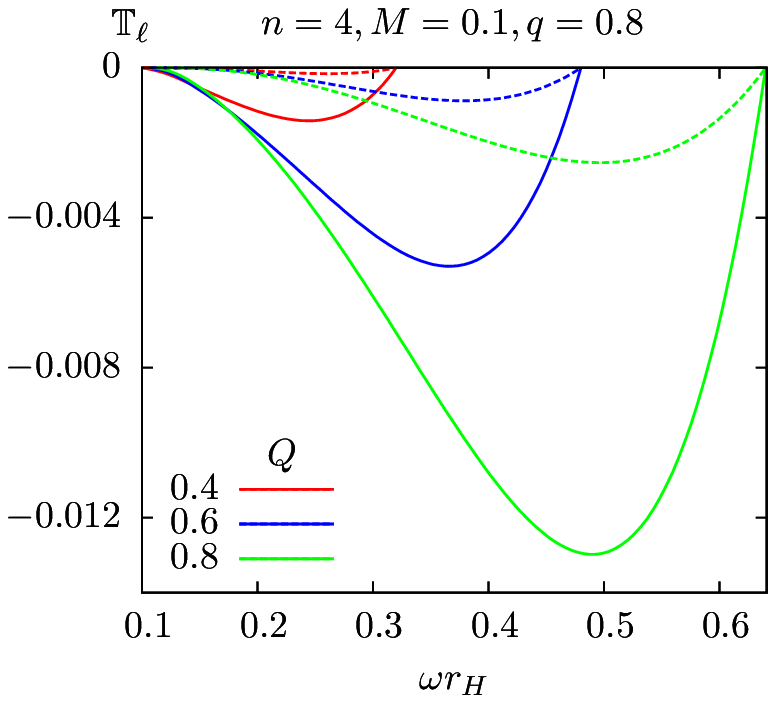}\hspace{2mm}
\includegraphics[clip=true,width=0.33\textwidth]{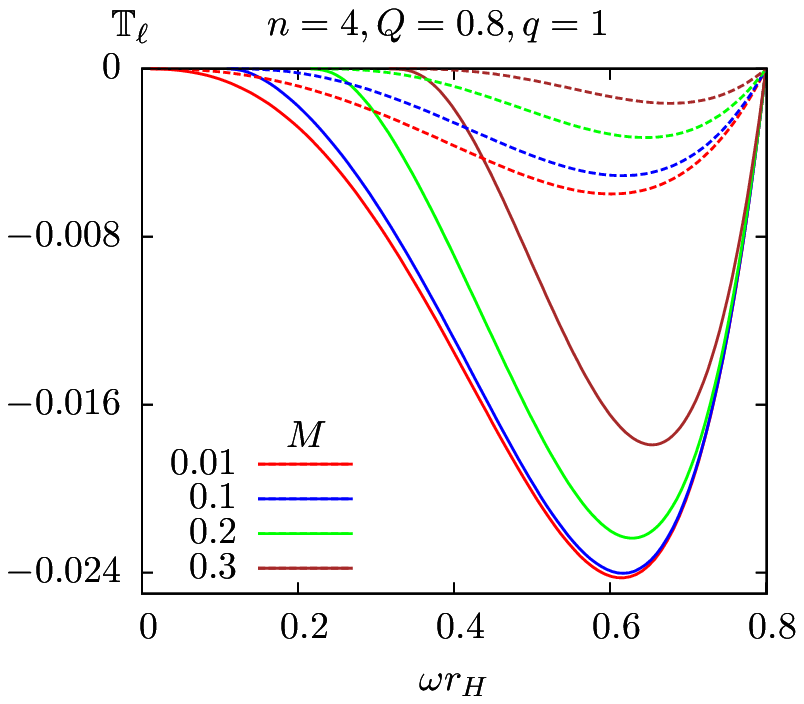}
\end{tabular}
\end{center}
\caption{\label{Superradiance01} Negative transmission factors that show the superradiant amplification of the coupled modes with $\ell= 1$ for different parameters. The two coupled modes are represented with the same colour (solid and dashed lines). The top and bottom rows differ in the space-time dimension. (Left panel)  Variation with the field charge $q$; (Middle panel) variation with the background charge $Q$; (Right panel) variation with the field mass.}
\end{figure*}

In Fig.~\ref{Superradiance01} the transmission factors for different masses, space-time dimensions and charges are illustrated to exhibit the superradiance phenomenon. As mentioned in the introduction, superradiant amplification of a bosonic field in a charged and/or rotating black hole occurs since there is Coulomb and/or rotational energy that can be extracted without decreasing the black hole area. The general condition of superradiance is $\omega<m\Omega_H+q\Phi_H$; when this condition is verified the transmission factor becomes negative and the scattered mode is amplified. In order to make the results clearer and more readable, here we just show the two coupled modes with $\ell=1$, as an example. Other cases are qualitatively similar. In the left panel of Fig.~\ref{Superradiance01}, we show the superradiance dependence on the field charge for $n=2$ (top row) and $n=4$ (bottom row). It is clear that superradiant amplification is enhanced with growing field charge except in the small energy regime; in this regime one observes, at least for the cases plotted, that the amplification decreases with increasing field charge (this effect is more noticeable for higher dimensions - bottom row). The middle panel of Fig.~\ref{Superradiance01} shows the dependence on the background charge - the trend is the same as when the field charge is varied, i.e. a generic superradiance enhancement with the charge except for small energies. In the right panel of Fig~\ref{Superradiance01}, we show the superradiance  dependence on the field mass; a suppression effect with increasing mass is observed, which is a generic behaviour for the transmission factor (independently of the spin of the field or of being in the superradiant regime)~\cite{Sampaio:2009ra,Sampaio:2009tp}.

After we obtain the transmission factors, the Hawking fluxes (\ref{eq:HawkFlux}) can be calculated directly. In the remainder we have chosen to show the flux of the number of particles. The flux of energy has similar features and can be simply calculated through multiplying each point by $\omega$.

\begin{figure*}
\begin{center}
\begin{tabular}{cc}
\includegraphics[clip=true,width=0.338\textwidth]{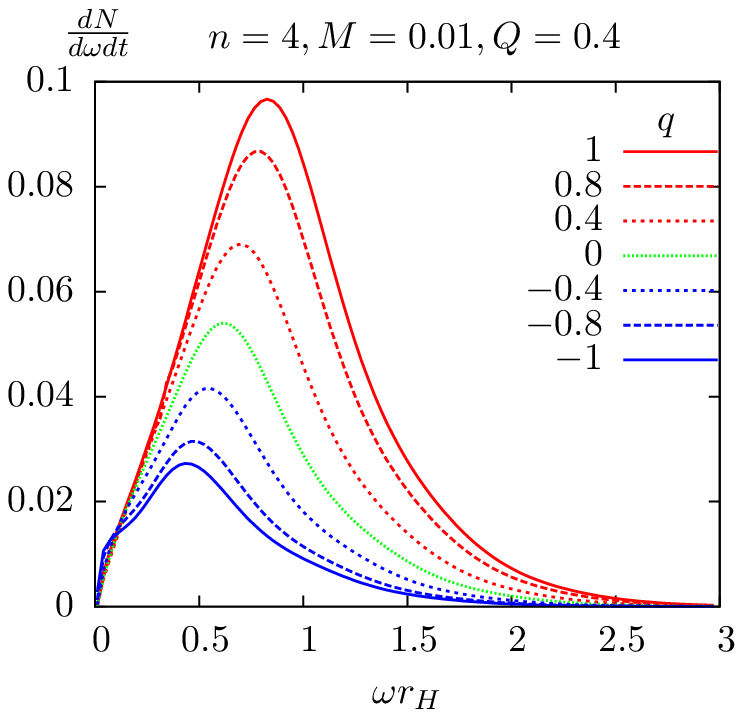}
\includegraphics[clip=true,width=0.45\textwidth]{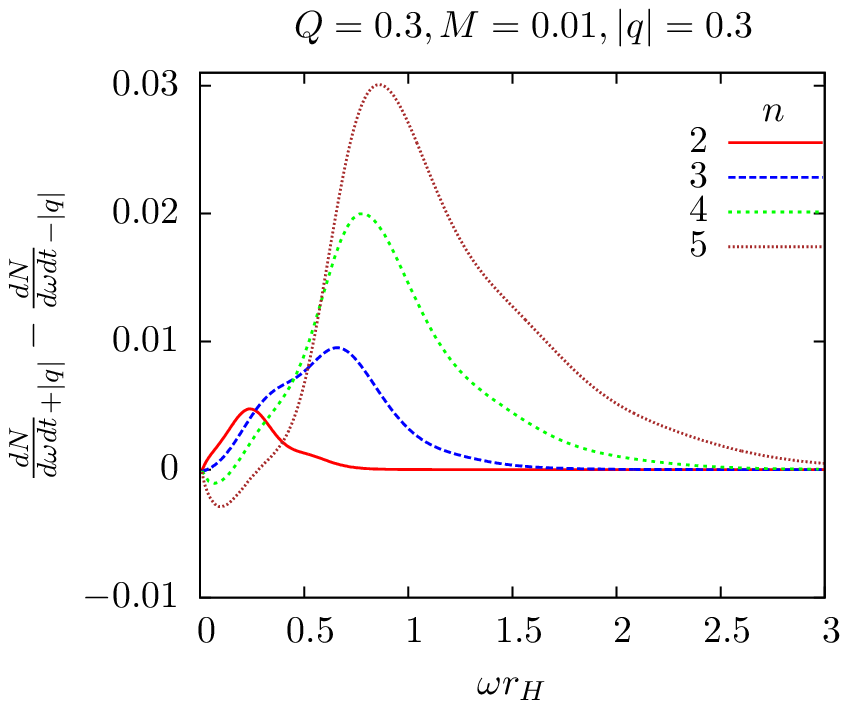}
\end{tabular}
\end{center}
\caption{\label{asymmetryeffect} (Left panel) Variation of the number fluxes with field charge for fixed $n$ and black hole parameters in the small field mass limit. (Right panel) Variation of the difference between positive charge and negative charge flux with $n$.}
\end{figure*}

In Fig~\ref{asymmetryeffect} we illustrate the number fluxes dependence on the field charge. On the left plots, we have kept the background charge parameter fixed (and positive), with $Q=0.4$ and $n=4$, varying the field charge $q$ from $-1$ to $1$. The plots show that there is a region at low energy where the negative field charge flux is larger, whereas in the remaining spectrum positive charge emission is favoured. It is also clear that if we integrate over the curves the emission of positive charge is always favoured. This low energy behaviour where negative charges are favoured, only occurs for $n>3$ as can be seen on the right panel, where the difference between the positive charge and negative charge flux spectrum is presented for various $n$. This inverted charge splitting effect was also observed for scalars and fermions~\cite{Sampaio:2009tp}, and it results from the interplay between the thermal factor and the transmission factor appearing in the expression for the number fluxes. Whereas the thermal factor always favours same charge emission, the transmission factor favours opposite charge emission and these factors dominate different parts of the spectrum. We have considered the same parameters as in~\cite{Sampaio:2009tp}, $Q=|q|=0.3$, wherein scalars and fermions have been studied, to allow for an easy comparison.

\begin{figure*}
\begin{center}
\begin{tabular}{ccc}
\hspace{-4.5mm}\includegraphics[clip=true,width=0.32\textwidth]{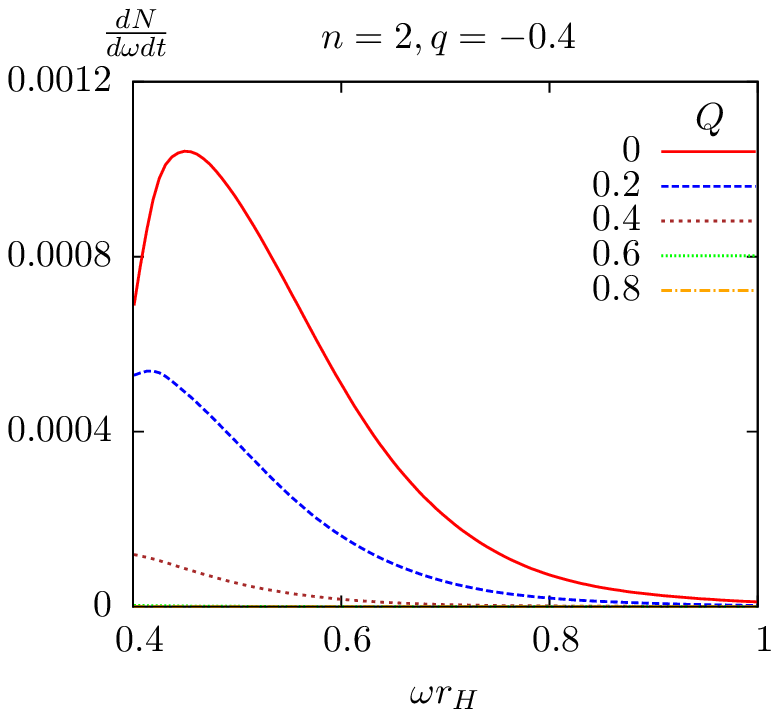} & \hspace{-4.5mm}
\includegraphics[clip=true,width=0.32\textwidth]{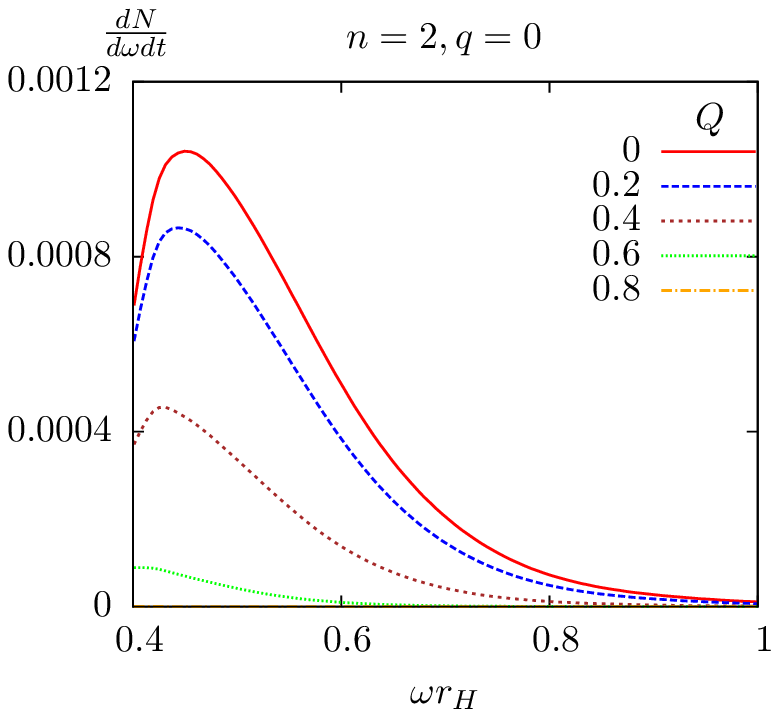} & \hspace{-5mm}
\includegraphics[clip=true,width=0.32\textwidth]{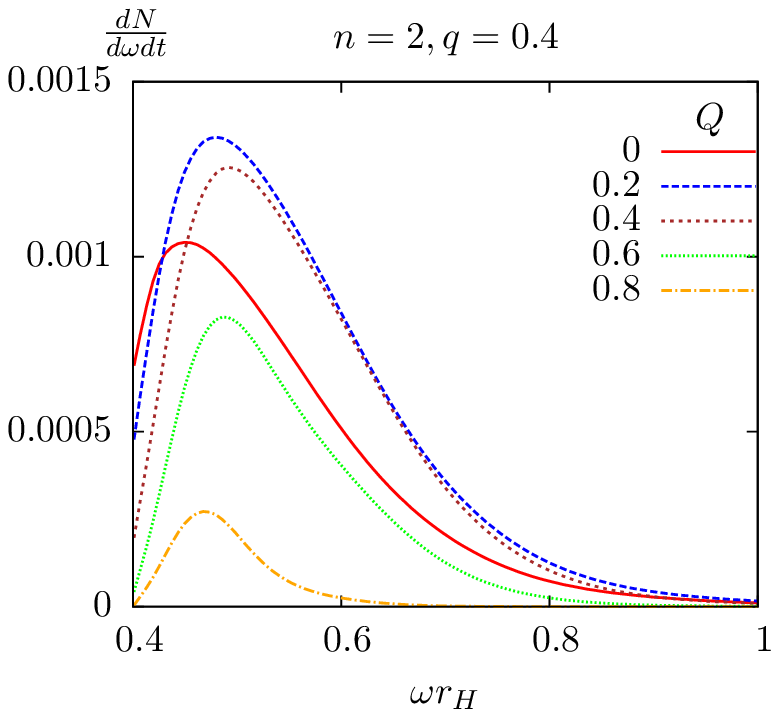}
\\
\hspace{-2mm}\includegraphics[clip=true,width=0.323\textwidth]{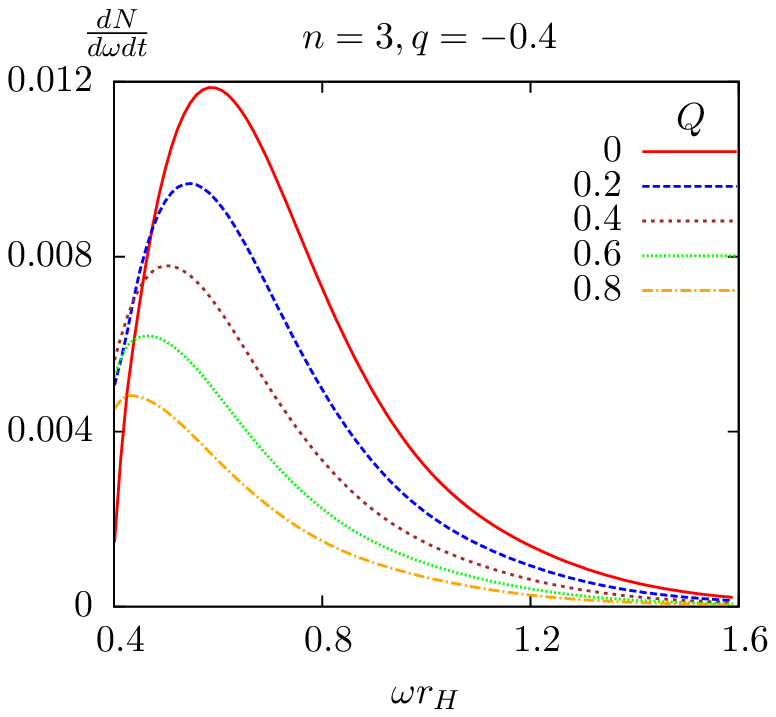} &
\hspace{-1mm}\includegraphics[clip=true,width=0.324\textwidth]{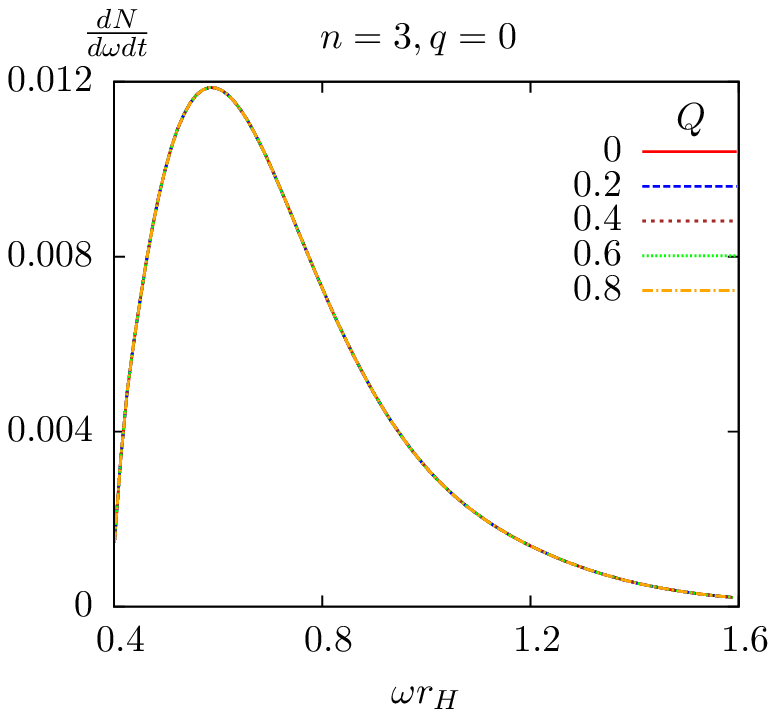} &
\hspace{-3mm}\includegraphics[clip=true,width=0.318\textwidth]{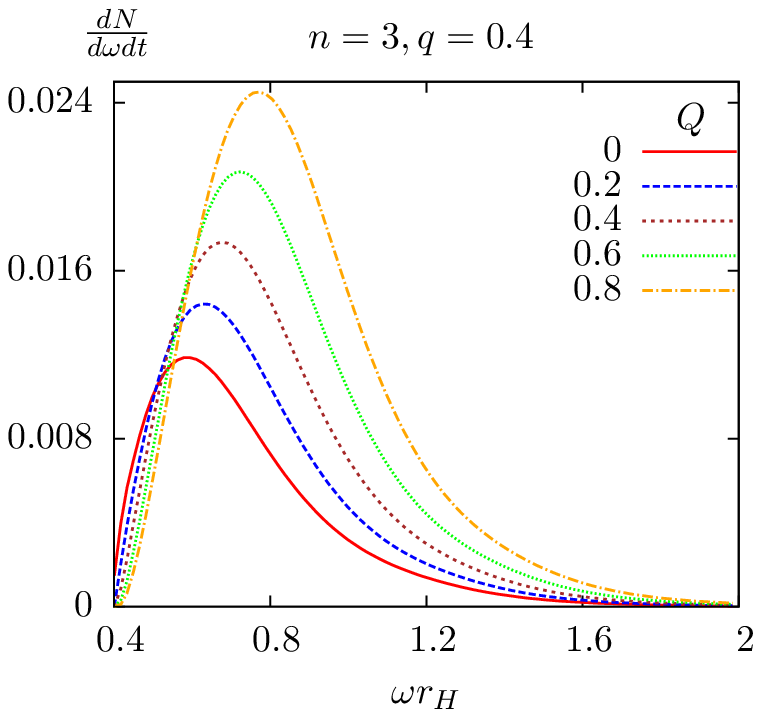}
\\
\includegraphics[clip=true,width=0.32\textwidth]{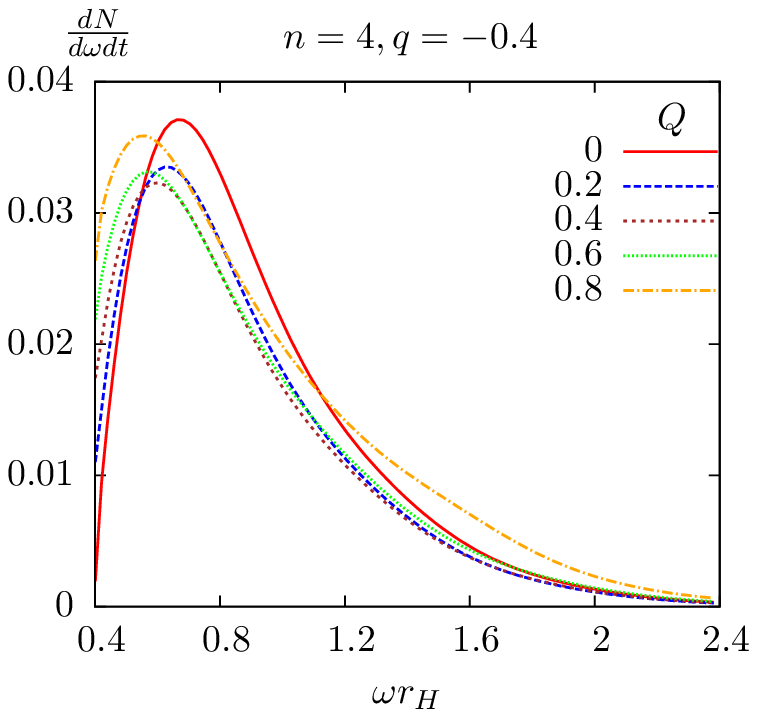} &
\includegraphics[clip=false,width=0.32\textwidth]{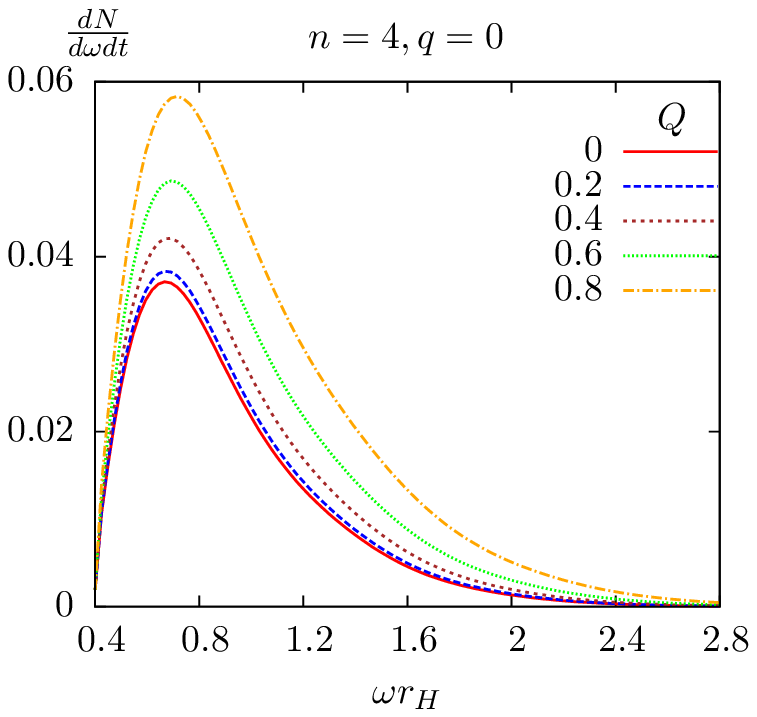} &
\includegraphics[clip=true,width=0.32\textwidth]{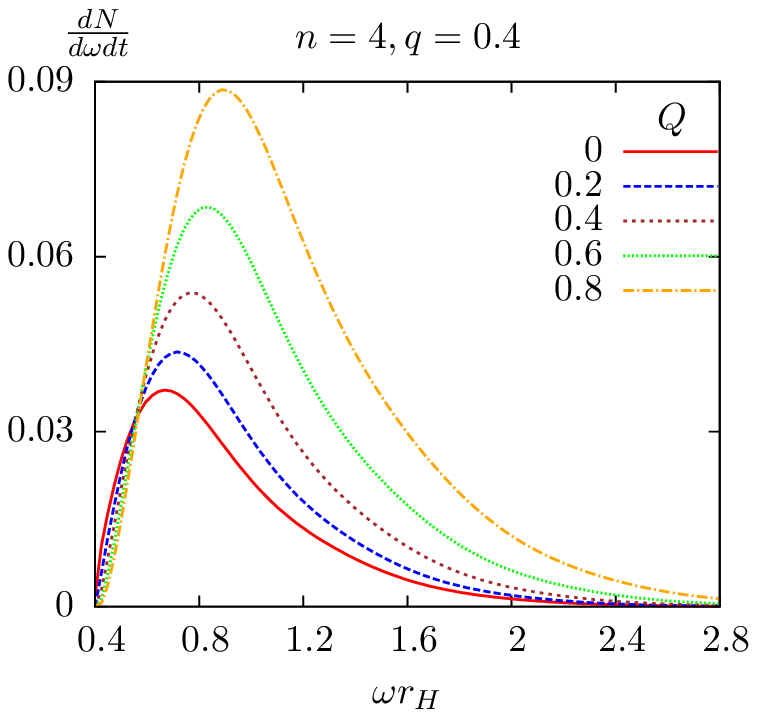}
\end{tabular}
\end{center}
\caption{\label{FluxVarQ} Number fluxes dependence on the background charge for different space-time dimensions and field charge, with fixed field mass $M=0.4$. We vary the field charge in the same row and vary the space-time dimension in the same column.
}\end{figure*}

In Fig.~\ref{FluxVarQ}, we present the number flux dependence on the background charge for different field charge in the same row and different space-time dimensions in the same column. Consider first the $q=0$ case (middle column); it shows that the fluxes are suppressed/maintained/enhanced with the increase of background charge for $n=2/n=3/n=4$. To understand this behaviour observe, from the definition of Hawking temperature, Eq.~\eqref{hawkingtemperature},  that the Hawking temperature is decreased/maintained/increased with increasing background charge for $n=2/n=3/n=4$, in these units. Since we are using horizon radius units, as we vary the charge parameter $Q$, we are actually varying the mass of the black hole as well as the charge while keeping $r_H=1$. Nevertheless, it is easy to see that, up to a stretching of the horizontal axis, if we fix the black hole mass and vary the dimensionful charge, these conclusions for the variation of the height of the curves do not change since the number flux is dimensionless~\footnote{The integrated flux however will not be the same for all background charges as expected, scaling as $r_H^{-1}$ for fixed black hole mass and varying charge.}. For higher temperature, one expects a larger flux of particles, which is indeed the behaviour shown in the second column of Fig.~\ref{FluxVarQ}. Turning on the field charge we observe a more involved behaviour. For $n=3$, in which the Hawking temperature does not vary with $Q$, we see in the first/third column and for sufficiently large energies a monotonic suppression/enhancement of the Hawking flux when the black hole has the opposite/same charge as the field. This is in agreement with the discussion of the left panel of Figure~\ref{asymmetryeffect}. For $n=2,4$, varying $Q$ one also varies the Hawking temperature and more complex patterns are observed.  Another trend is that the number fluxes increase as the space-time dimension increases, which may be understood from the existence of more modes that contribute to the transmission factor.

\subsection{Mass effect and bulk/brane emission}

\begin{figure*}
\begin{center}
\begin{tabular}{ccc}
\includegraphics[clip=true,width=0.33\textwidth]{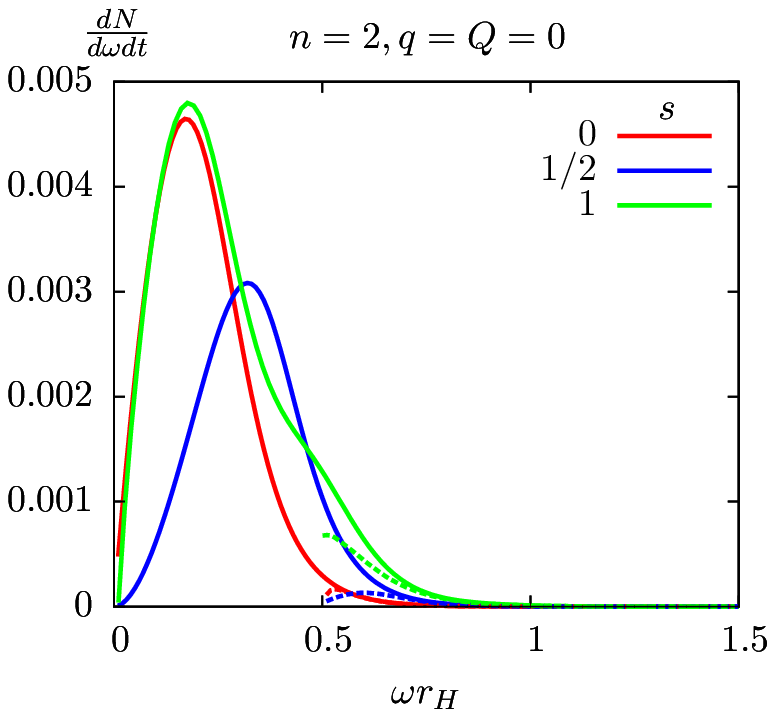}
\includegraphics[clip=true,width=0.331\textwidth]{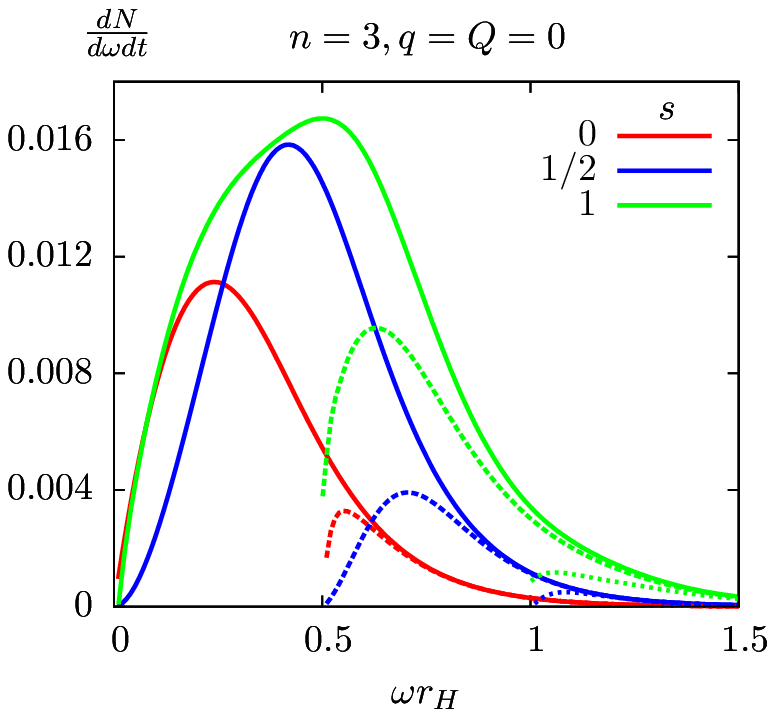}
\includegraphics[clip=true,width=0.324\textwidth]{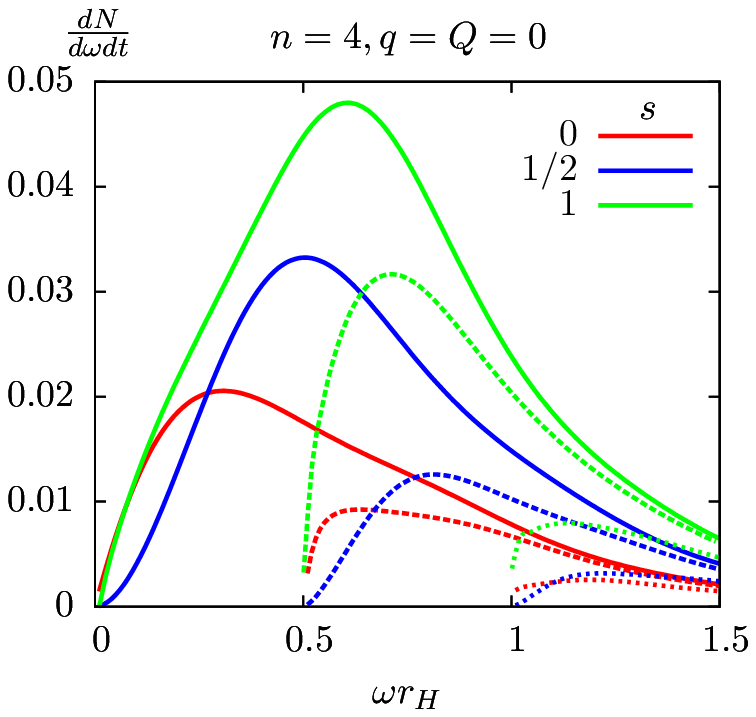}
\end{tabular}
\end{center}
\caption{\label{Mass_spin} Variation of the number fluxes for neutral particles with different spins (scalar, fermion and Proca) on the brane. For each type of particles we consider three different masses ($M=0$, $0.5$ and $1$), which can be identified by the starting point of the curve. Note we have included the two helicities for the fermion field and all the three modes for the Proca field (we have used a small mass $M=0.01$ for the latter, instead of $M=0$).}
\end{figure*}
In Figure~\ref{Mass_spin}, we perform a comparison of the effect of introducing a mass term for the various spins which are relevant for the Standard Model brane degrees of freedom. We have used the data of~\cite{Sampaio:2009tp} for scalars and fermions. Note that for fermions we have multiplied the data by a factor of two to take into account the two helicities of the Dirac field, since here we are also considering all the three modes for the Proca field. The dashed curves are for increasingly larger field mass (we have used the cases $M=0$, $0.5$ and $1$, as can be seen from the threshold points where the curves start). For $D=n+2=4$ (left panel), we observe a striking similarity between the Proca ($s=1$) flux with the scalar ($s=0$) flux for $M=0$ (except at the high energy tail), which is due to the dominance of the $\ell=0$ mode. We can see this feature is always true at small energies for larger $n$ (middle and right panels);  as we increase $n$, however, higher modes of the Proca field enhance the flux as compared to scalars. The extra modes of the Proca field contributing at higher energy, also explain the fact that the mass suppression is not as large as for Dirac fermions or scalars, as we see from the dashed curves corresponding to $M=0.5$ for example. Contrasting with the low energy behaviour, where scalars and Proca field are dominated by the $s$-wave, whereas Dirac fermions are suppressed since they do not allow an $s$-wave, is the high energy behaviour. There we observe tails which are in the ratio $1:2:3$ following the number of degrees of freedom for the scalar, fermion and Proca fields respectively. This is in agreement with the fact that all transmission factors tend to one at high energy.

In the remainder, we combine our results for the Proca field on the brane with those in~\cite{HSW:2012} for a Proca field in the bulk, to analyse the relative bulk-to-brane emissivity in the neutral case.

\begin{table}[h]
\begin{center}
\resizebox{3.0in}{0.38in}{
\begin{tabular}{|c|c|c|c|c|c|c|}
\hline
 & $n=2$ & $n=3$ & $n=4$ & $n=5$ & $n=6$ & $n=7$
\\
\hline
$M=0.1$ & 1 & 0.46 & 0.38 & 0.41 & 0.53 & 0.83
\\
\hline
$M=0.3$ & 1 & 0.49 & 0.40 & 0.42 & 0.55 & 0.86
\\
\hline
$M=0.6$ & 1 & 0.59 & 0.47 & 0.49 & 0.62 & 0.95
\\
\hline
\end{tabular}
}
\end{center}
\caption{Bulk-to-Brane relative energy emission rates for massive neutral vector fields for different mass $M$ in terms of the space-time dimension $D=2+n$.}
\label{bulkbranecomparison}
\end{table}

The total energy rate (or number rate) emitted in Hawking radiation for a given field, is given by integrating the fluxes of Eq.~\eqref{eq:HawkFlux} (or their counterparts in the bulk) over $\omega$. The bulk-to-brane energy emissivity ratio, for massive neutral Proca fields of different mass is shown in Table~\ref{bulkbranecomparison} as a function of $n$. We have used the data in~\cite{HSW:2012} to obtain the total emissivity for bulk fields, as well as the data presented here, for the brane emissivity. The entries of Table~\ref{bulkbranecomparison}, show clearly that the emission of energy into brane-localised Proca particles is dominant, for $n$ larger than two, which is consistent with the argument that black holes radiate mainly on the brane~\cite{EHMprl2000}. For fixed mass, as $n$ increases, the bulk-to-brane energy emission ratio initially decreases, reaching a minimum value for an intermediate $n$. However, if one increases $n$ further, the bulk-to-brane energy emission ratio increases again. Furthermore, we have also observed that the bulk-to-brane energy emission ratio increases with the field mass. A similar behaviour  was observed for scalar fields~\cite{Harris:2003eg,Kanti-Pappas2010}. Finally, we found that the bulk-to-brane energy emission rate for the Proca field is larger than that for the scalar field for all $n$, being more noticeable for large $n$, say $n=6$ and $n=7$ (note the different notation for $D=2+n$ in our paper as compared with~\cite{Harris:2003eg,Kanti-Pappas2010}).

\section{Discussion and Final Remarks}
\label{discussion}
In this paper we have completed the analysis started in~\cite{HSW:2012}, by  computing the transmission factor for a charged Proca field propagating in the background of a charged black hole on a brane. Furthermore, this study completes the study of the effect of mass and charge for particles evaporating on the brane, for all spins relevant for the Standard Model in brane world scenarios~\cite{Sampaio:2009ra,Sampaio:2009tp}. Since the Proca field equations for the various modes do not totally decouple, we have followed a numerical strategy, to determine the transmission factors as in~\cite{HSW:2012}.

One of the main, and novel features arising from considering a charged Proca field on a charged background is the existence of negative transmission factors, a signature of superradiant scattering, which we have presented and described. Our results are consistent with the condition of superradiance as allowed by the area theorem, and we found that generically increasing the field or background charge amplifies the effect (though some inversions are possible at small energies). It is worth commenting that although superradiant scattering is observed for both rotating and charged black holes, rotating black holes exhibit superradiant instabilities against massive scalar fields, and in particular against the Proca field~\cite{paniPRL,paniPRD}, whereas charged (non-rotating) black holes do not seem to exhibit  analogous instabilities against charged, massive bosonic fields. This can be checked for the Proca field using our numerical framework, by computing the frequencies of quasi-bound states, a study that can also teach us how long lived Proca hair around charged black holes can be. Thus, even though our setup offers the technical advantage of spherical symmetry for studying superradiant scattering of the Proca field, it cannot be used for studying superradiant instabilities, and in particular their non-linear development. Given the lack of spherical symmetry in the rotating case, such study will require sophisticated numerical relativity methods of the sort used in~\cite{Witek:2010qc,Dolan:2012yt,Witek:2012tr}.

An effect observed herein,is that similarly to scalars and fermions, there is an inverted charge splitting effect at small energies for more than one extra space-time dimension, where particles with charge opposite to the black hole are emitted dominantly. Nevertheless, for larger energies, the normal splitting order, favouring the emission of same charge particles as to discharge the black hole is restored and overall (by integrating out the flux), this channel tends to discharge the black hole. This effect has been suggested to be another signature that could be found in the energy spectrum of charged fermions \cite{Sampaio:2009tp}. Since fermions in the final state are also produced indirectly from the decay of vector bosons such as $W^{\pm}$ or the $Z$ particles one may ask whether the effect survives. Here we have verified that the effect is present for $W^{\pm}$, so at least for the case when the final state decay products are a charged lepton $\ell^\pm$ and a neutrino, this contributions will certainly enhance the effect.

In the neutral case, we have performed two comparisons. First we have compared the effect of the mass and spin on the Hawking fluxes, for all spins in the Standard model. Our main findings are that the Proca field spectrum departs from being similar to a scalar field in four dimensions and becomes increasingly dominant for larger number of dimensions, peaking and extending towards larger energies. This also means that the mass suppression effect is smaller for the Proca field than it is for scalars and fermions. Secondly, we have compared the bulk-to-brane emission ratio for Proca fields, confirming brane dominance in general, and the suppression of brane dominance with increasing mass.

Our methods and results can be used to improve the modelling of black hole evaporation in TeV gravity scenarios, in the black hole event generators~\cite{Frost:2009cf,Dai:2007ki} that are in use at the ATLAS and CMS experiments to put bounds on extra dimensions in this channel~\cite{CMS:2012yf,ATLAS-CONF-2011-065,ATLAS-CONF-2011-068,Gingrich:2012vs}.

\bigskip

\noindent{\bf{\em Acknowledgements.}}
 M.W. and M.S. are funded by FCT through the grants SFRH/BD/51648/2011 and SFRH/BPD/69971/2010. The work in this paper is also supported by the grants PTDC/FIS/116625/2010 and  NRHEP--295189-FP7-PEOPLE-2011-IRSES.
\appendix

\section{Functions and matrices}
The functions that are used in the text are (where $\kappa_0^2=\ell(\ell+1)$):
\begin{equation}
A(r)\equiv \sum^{2n+1}_{m=0}{a_m y^m}=r \left[r^n-(1+Q^2)r+Q^2r^{n-2}\right]^2 \; ,\nonumber
\end{equation}
\begin{equation}
B(r)\equiv \sum^{2n}_{m=0}{b_m y^m}=2\left[r^n-(1+Q^2)r+Q^2r^{n-2}\right]^2\; ,\nonumber
\end{equation}
\begin{multline}
C(r)\equiv \sum^{2n+1}_{m=0}{c_m y^m}= (\omega r-qQ)^2r^{2n-1}\\
-(\kappa_0^2+M^2r^2)r^{n-1}\left[r^n-(1+Q^2)r+Q^2r^{n-2}\right]  \; ,\nonumber\\
\end{multline}
\begin{multline}
E(r)\equiv \sum^{2n}_{m=0}{e_m y^m}=iqQr^{n-1}\left[(1+Q^2)(n-1)r\right. \\ \left.-2Q^2r^{n-2}\right]+ i\omega r^n\left[2r^{n}-(1+Q^2)(n+1)r+4Q^2r^{n-2}\right] \, ,\nonumber
\end{multline}
\begin{equation}
\tilde A(r)\equiv \sum^{2n}_{m=0}{\tilde a_m y^m}=\left[r^n-(1+Q^2)r+Q^2r^{n-2}\right]^2  \; ,\nonumber
\end{equation}
\begin{equation}
\tilde B(r)\equiv 0 \;,\nonumber
\end{equation}
\begin{multline}
\tilde C(r)\equiv \sum^{2n}_{m=0}{\tilde c_m y^m}= (\omega r-qQ)^2r^{2n-2}\\
 -\left(\kappa_0^2+M^2r^2\right)r^{n-2}\left[r^n-(1+Q^2)r+Q^2r^{n-2}\right] ,\nonumber
\end{multline}
\begin{multline}
\tilde E(r)\equiv \sum^{2n-2}_{m=0}{\tilde e_m y^m}=iqQ r^{n-1}\left[(n-3)(1+Q^2)+\right.
\nonumber \\
 \left.+2r^{n-1}\right] -i\omega r^{n-1}\left[(1+Q^2)(n-1)r-2Q^2r^{n-2}\right]\;.\nonumber
\end{multline}
The recurrence relations are
\begin{eqnarray}
\mu_0&=&\nu_0\;,\nonumber\\
\mu_1&=&-\frac{\left[a_3\rho(\rho-1)+b_2\rho +c_1+e_1\right]\nu_0+ e_0 \nu_1}{a_2\rho(\rho+1)+c_0}\;,\nonumber
\end{eqnarray}
\begin{eqnarray}
\mu_j&=&\frac{\tilde a_2(\rho+j)(\rho+j-1)+\tilde c_0}{D_j}f_j-\frac{e_0}{D_j}\tilde
f_j\;,\nonumber\\
\nu_j&=&\frac{a_2(\rho+j)(\rho+j-1)+c_0}{D_j}\tilde
f_j-\frac{\tilde e_0}{D_j}f_j\;,
\label{recurone}
\end{eqnarray}
with
\begin{multline}
D_j=\Big[a_2(\rho+j)(\rho+j-1)+c_0\Big]\times \\
\times \left[\tilde a_2(\rho+j)(\rho+j-1)+\tilde c_0\right]-\tilde e_0 e_0\;,\nonumber
\end{multline}
\begin{multline}
f_j=-\sum^j_{m=1}\left[(a_{m+2}(\rho+j-m)(\rho+j-m-1)\right. \\ \left.  +b_{m+1}(\rho+j-m)+c_m\big)\mu_{j-m}+e_m\nu_{j-m}\right]\;,\nonumber
\end{multline}
\begin{multline}
\tilde f_j=-\sum^j_{m=1}\left[\big(\tilde a_{m+2}(\rho+j-m)(\rho+j-m-1)+\tilde c_m\big) \right.\\
\left. \times \nu_{j-m}+\tilde e_m\mu_{j-m}\right]\;.\nonumber
\end{multline}
\begin{widetext}
The coefficients used in the asymptotic expansion in the text are
\begin{eqnarray}
c^\pm&=&\dfrac{i}{2\omega}\Big[-\kappa_0^2+Q^2(M^2-2\omega^2)-\dfrac{q^2Q^2M^2}{k^2}\mp i\dfrac{qQ\omega}{k}+(2\omega^2-M^2)(1+Q^2)\delta_{3,n}\nonumber\\&+&\Big(\pm i\dfrac{M^2(1+Q^2)}{2k}-\Big(\dfrac{M^2(1+Q^2)}{2k}\Big)^2+(1+Q^2)^2(2\omega^2-M^2)+\dfrac{qQ\omega(1+Q^2)(M^2-2k^2)}{k^2}\Big)\delta_{2,n}\Big] \; .\nonumber\\\label{cpm}
\end{eqnarray}
The relation between the new first order radial functions $\mathbf{\Psi}$ used in~\eqref{eq:ODEcoupled}, and the 4-vector $\mathbf{V}^T=(\psi,d_r\psi,\chi,d_r\chi)$ for the original fields and derivatives is found from Eqs.~\eqref{asymptoticpsi},~\eqref{asymptoticpchi} and their derivatives. The corresponding $r$-dependent matrix transformation $\mathbf{T}$ is defined
\begin{equation}
\mathbf{V}= \mathbf{T} \mathbf{\Psi} \; ,
\end{equation}
 and its form is
\begin{equation}\label{eq:T}
\mathbf{T}=\left(\begin{array}{cccc}\frac{e^{i\Phi}}{r} & \frac{e^{-i\Phi}}{r} & e^{i\Phi} & e^{-i\Phi} \vspace{2mm}\\ \frac{ike^{i\Phi}}{r} & -\frac{ike^{-i\Phi}}{r} & \left[ik+\frac{i\varphi}{r}\right]e^{i\Phi} & -\left[ik+\frac{i\varphi}{r}\right]e^{-i\Phi} \vspace{2mm}\\ 0 & 0 & \left(-\frac{k}{\omega}+\frac{c^+}{r}\right)e^{i\Phi}  & \left(\frac{k}{\omega}+\frac{c^-}{r}\right)e^{-i\Phi} \vspace{2mm}\\ 0 & 0 & \left[-\frac{ik^2}{\omega}+\frac{ikc^+-\frac{ik\varphi}{\omega}}{r}\right]e^{i\Phi}& -\left[\frac{ik^2}{\omega}+\frac{ikc^-+\frac{ik\varphi}{\omega}}{r}\right]e^{-i\Phi} \end{array}\right) \ ,
\end{equation}
On another hand, the original system~\eqref{coupledequations}, can be written in a first order form
\begin{equation}
\dfrac{d\mathbf{V}}{dr}=\mathbf{X}\mathbf{V} \; ,
\end{equation}
where the matrix $\mathbf{X}$ is
\begin{equation}\label{eq:X}
\mathbf{X}=\left(\begin{array}{cccc}0 & 1 & 0 & 0 \vspace{2mm} \\ -\dfrac{C}{A} & -\dfrac{B}{A}& -\dfrac{E}{A}  & 0 \vspace{2mm}\\0 & 0 & 0 & 1 \vspace{2mm} \\ -\dfrac{\tilde E}{\tilde A}  & 0& -\dfrac{\tilde C}{\tilde A}& 0 \end{array}\right) \ ,
\end{equation}
and the 2-vectors
\begin{equation}\label{eq:yplus}
\mathbf{y}^{\pm}=\left(\begin{array}{c}\sqrt{\dfrac{\kappa_0^2 k}{\omega^2}} a_0^{\pm} \\ i\sqrt{\dfrac{k}{M^2}}\left[\left(\pm \varphi+\omega c^{\pm}\mp k(1+Q^2)\delta_{n,2}\pm\dfrac{kqQ}{\omega}\right)a_0^{\pm}\pm ka_1^{\pm}\right] \end{array}\right) \ ,
\end{equation}
\begin{equation}\label{eq:hminus}
\mathbf{h}^-=\left(\begin{array}{c}\sqrt{t} \nu_0 \\ \sqrt{\dfrac{M^2}{M^4+q^2Q^2}}\left(-i\kappa_0^2\nu_0+(\omega-qQ)d \right) \end{array}\right) \; ,
\end{equation}
with
\begin{equation}
t=\dfrac{\kappa_0^2}{M^4+q^2Q^2}\Big[(\omega-qQ)\left(qQa+bM^2+2i\rho M^2\alpha+\dfrac{2qQM^2}{\beta}\right)-\kappa_0^2M^2\Big]\;,\nonumber
\end{equation}
\begin{equation}
d=\left[\left(1+\rho-\dfrac{iqQ}{M^2}\rho\right)(a+ib)+\left(1+\dfrac{iqQ}{M^2}\right)\left(\dfrac{iqQ}{\beta}-\alpha\rho\right)\right]\nu_0+(1+2\rho)\left(\dfrac{iqQ}{M^2}-1\right)\nu_1 \;,\nonumber
\end{equation}
\begin{equation}
a=\dfrac{qQ}{\omega-qQ}\;,\nonumber
\end{equation}
\begin{equation}
b=\dfrac{qQ(2+n(n-3)(1+Q^2))}{\beta^2}-\dfrac{2\omega\alpha}{\beta}+\dfrac{\kappa_0^2+M^2}{\omega-qQ}\;,\nonumber
\end{equation}
\begin{equation}
\alpha=\dfrac{n(n-1)+(n-3)(n+2)Q^2}{2\beta}\;,\nonumber
\end{equation}
\begin{equation}
\beta=(n-1)+(n-3)Q^2\;.\nonumber
\end{equation}
\end{widetext}

\bibliographystyle{h-physrev4}
\bibliography{proca-charged}


\end{document}